\pdfoutput=1
\documentclass{sig-alternate-10pt}
\usepackage{times}
\usepackage{balance}
\usepackage{multirow}
\usepackage{epsfig}
\usepackage{graphicx}
\usepackage{subfigure}
\usepackage{cite}
\usepackage{hyperref}
\usepackage{cleveref}
\usepackage{textcomp}
\usepackage{listings}
\usepackage{url}

\usepackage{subfigure}
\usepackage[utf8]{inputenc}
\usepackage[T1]{fontenc}
\usepackage{cases}
\usepackage{amsmath} 
\usepackage{amssymb} 
\usepackage{amsfonts}
\usepackage{color}
\usepackage{tabu}
\usepackage{pifont}

\newenvironment{packed_itemize}{
\begin{list}{\labelitemi}{\leftmargin=1.2em}
\setlength{\itemsep}{1pt}
\setlength{\parskip}{0pt}
\setlength{\parsep}{0pt}
\setlength{\headsep}{0pt}
\setlength{\topskip}{0pt}
\setlength{\topmargin}{0pt}
\setlength{\topsep}{0pt}
\setlength{\partopsep}{0pt}
}{\end{list}}

\newfont{\mycrnotice}{ptmr8t at 7pt}
\newfont{\myconfname}{ptmri8t at 7pt}
\let\crnotice\mycrnotice%
\let\confname\myconfname%

\newcommand{\para}[1]{{\vspace{3pt} \bf \noindent #1 \hspace{4pt}}}
\begin{document}
\title{Empirical Analysis of Password Reuse and Modification across Online Service}

 \author{Chun Wang, Steve T.K. Jan, Hang Hu, Gang Wang\\
 \affaddr{Department of Computer Science, Virginia Tech}\\
 {\em \{wchun, tekang, hanghu, gangwang\}@vt.edu}}
\maketitle

\begin{abstract}
Leaked passwords from data breaches can pose a serious threat to
users if the password is reused elsewhere. With more online services getting breached today, there is still a lack of
large-scale quantitative understanding of the risks of password reuse across services.
In this paper, we analyze a large collection of 28.8 million users and 
their 61.5 million passwords across 107 services. We find that 38\% of
the users have reused exactly the same password across different
sites, while 20\% have modified an existing password to create new
ones. In addition, we find that the password modification patterns are
highly consistent across different user demographics, indicating a
high predictability. To quantify the risk, we build a new
training-based guessing algorithm, and show that more than 16 million
password pairs can be cracked within just 10 attempts (30\% of the modified passwords and all
the reused passwords). 
\end{abstract}


\section{Introduction}
The widespread of data breaches ({\em e.g.}, Yahoo, Myspace,
Office of Personnel Management, Ashley Madison) are posing
significant threats to users and organizations. In
2016 alone, there were more than 2000 confirmed
breaches causing a leakage of billions of user
records~\cite{verizon17}. Many of the leaked datasets contain
sensitive information such as {\em user passwords}, which are often made publicly available on the Internet by
attackers~\cite{new11,new10,new8,new7,new9}. 

A leaked password incurs serious risks if the user reuses the password
across different services. Reusing the same or even slightly
modified passwords allows attackers to further compromise the user's accounts in
other ``un-breached'' services~\cite{new13,new14}. Even worse, if the
target user happens to be the administrator of another service,
password reuse may lead to new massive data breaches ({\em e.g.},
Dropbox~\cite{new12}).


With more passwords leaked everyday~\cite{site1,site5}, there is still a
lack of large-scale quantitative understanding on password usage
across online services. Existing work studies password
reuse and transformation either through a user survey~\cite{Komanduri:2011,DasBCBW14}
or small-scale data analytics (6--7K
users)~\cite{Zhang:2010:,DasBCBW14}. The limited scope of the data
(sample size, service type, user demographics) makes it challenging to
comprehensively quantify the risks to see the bigger picture.

In this paper, we conduct a large-scale measurement on 28,836,775
users on their password reuse and modification patterns
across 107 online services.\footnote{Our study has received IRB
  approval (Protocol \#17-393).} 
By analyzing publicly available password datasets, we seek to
empirically measure the common ways in which users reuse/modify
passwords. In addition, to quantify the security risks introduced by
password reuse and modification, we develop  a
{\em training-based} algorithm to guess a target user's
password based on her leaked one. Our study reveals several key findings.


\para{How often do users reuse or modify existing passwords? }\\
Among the 28.8 million users, we find that 38\% of users have
once reused the same password in two different services and 21\% of
the users once modified an existing password to sign up a new service.  
Particularly, passwords of email services ({\em e.g.}, Gmail) have a noticeably high reuse
rate (60.4\%). Given the sensitivity of email accounts, reusing
the password of email accounts incurs serious risks. 

\para{What are the common ways of modifying passwords?}\\
We empirically measure 8 high-level categories of password
transformation rules. We find that users prefer using simple rules
to modify passwords. More importantly, the password
transformation patterns are highly consistent across users of
different professions (military, government, education) and
countries. The low variance of the transformation patterns is
likely to make the modified passwords predictable.


\para{How likely can attackers guess a modified password?}\\
Our training-based algorithm can guess 30\% of the modified passwords
within 10 attempts (46.5\% within 100 attempts). Together with the
identical passwords (reused), more than 16 million
password pairs can be cracked within 10 guesses.  In addition, the
algorithm achieves a similar performance even if it is trained with
only 0.1\% of the data. This confirms the low-variance of password
modification patterns, indicating that attackers can learn the basic
patterns with minimal training to crack massive passwords in an {\em
  online fashion}.



This work is the first large-scale measurement on
the password reuse and modification patterns across online
services. Our result sheds light on the emerging
security threats introduced by
massive data breaches, and calls for more effective tools to secure
users' online accounts and digital assets.

\section{Related Work}
\label{sec:related}

\para{Password Reuse \& Transformation.}
Password guessing attacks become a major concern as data breaches are
increasingly frequent. The attack is immediately
effective if the user reuses {\em the same} password for
different services~\cite{www07pass,Bailey2014}. Even if non-identical
passwords are used, users may follow simple transformation patterns 
to modify their passwords, which makes their
passwords predicable~\cite{DasBCBW14,Zhang:2010:}. Existing work
investigated this problem based on
user surveys or small-scale data analytics (6--7K
users)~\cite{DasBCBW14,Zhang:2010:,Komanduri:2011,197316}. In this work, we
perform the first a large-scale measurement to understand cross-site password
usage and quantify the risk of leaked passwords (28.8 million users, 107 websites).


\para{Online \& Offline Password Guessing.} 
Online password guessing requires attackers to guess the password within a
limited number of attempts. Trawling based approach simply guesses
the most popular passwords chosen by users~\cite{Mazurek:2013:}.
More targeted guessing exploits the fact that users may reuse the
same/similar passwords across services~\cite{DasBCBW14,Zhang:2010:} or
include PII information (name, birthday) in their
passwords~\cite{Wang:2016:,info16:}. 

A larger body of work focuses on offline
guessing~\cite{pars15,ur2016,VerasCT14, Weir:2009:PCU,
  KelleyOakland2012, password15, Narayanan:2005}, where the number of
attempts is unlimited. A common scenario is that given a hashed
password dataset, offline guessing seeks to recover the
plaintext passwords. A number of approaches have been proposed,
including Markov Models~\cite{Narayanan:2005, Ma:2014:SPP:}, Mangled
Wordlist methods~\cite{ur2015}, Probabilistic
Context-Free Grammars
(PCFGs)~\cite{Weir:2009:PCU,KelleyOakland2012,VerasCT14,Narayanan:2005},
and Neural Networks~\cite{ur2016}. Offline guessing has also been used to
measure password strength~\cite{Ur:2012:YPM:,CarnavaletM14,pars15}.

\begin{table}[t]
\centering
\begin{small}
\begin{tabu}{l|l|l}
\tabucline[1.1pt]{-}
 Category & \#Plain PWs & Top 3 Largest Datasets \\
 & (\#Datasets) &  \\ 
\hline
Social & 286M (7)  &  Myspace, VK, LinkedIn  \\
Adult & 75.2M (9) & Zoosk,  Mate1, YouPorn  \\
Game & 40.8M (13) &   Neopets, 7k7k, Lbsg \\
Entertain  & 30.7M (4)  & Lastfm, Swingbrasileiro, LATimes \\
Internet  & 16.4M (18) & 000webhost, Comcast, Yahoo \\
Email & 9.6M (3) & Gmail, Mail.ru, Yandex \\
Forum & 1.1M (25)  & CrackingForum, Abusewith.us, Gawker \\
Shopping &  340K (12)  & RedBox, 1394store, Myaribags \\
Others &  210K (7)  & Data1, Data2, Data3 \\
Business & 10K (9) & Movatiathletic, Hrsupporten, 99Fame \\
\hline
Total & 460M (107) &  Myspace, VK, LinkedIn \\
\tabucline[1.1pt]{-}
\end{tabu}
\vspace{-0.12in}
\caption{Categories and statistics of collected datasets. }
\label{tab:cat}
\end{small}
\vspace{-0.12in}
\end{table}

\section{Dataset} 
\label{sec:data}
To analyze password usage across services, we gathered a
large collection of publicly available password datasets. In January
2017, we searched through online forums and public
data archives for candidate datasets using two criteria. First, the dataset
should contain email addresses to link a user's passwords
across services. Second, we exclude datasets with only {\em salted hashes} since
it is difficult to massively recover their passwords. 

\begin{figure}[t]
 \centering
\begin{minipage}{0.4\textwidth}
 \centering
	\includegraphics[width=1\textwidth]{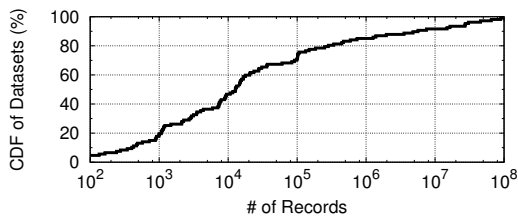}
	\vspace{-0.27in}
	\caption{\# of Passwords in each dataset.}
	\label{fig:size}
\end{minipage}
\vspace{-0.1in}
\end{figure}

We collected 107 datasets leaked between 2008--2016, which contain 497,789,976 passwords and
428,199,842 unique users (email addresses). 14 datasets contain hashed
passwords, and we spent a week to recover the
plaintext using offline guessing tools~\cite{JtR, ur2015,
  pars15}. In total, we obtained 460,874,306 plaintext passwords (93\% of all passwords). 
The rest 7\% are difficult to recover, and we will use them
to test our guessing algorithm later. Figure~\ref{fig:size} shows the
number of passwords in each dataset. In Table~\ref{tab:cat}, we
classify the datasets into 10 categories. The ``others'' category contains 7
datasets with generic file names (difficult to label). We have made sure that the 7
datasets did not overlap with any existing ones. 


\para{Primary Dataset (28.8 Million Users).} To study cross-site
password usage, we need users who appear on at least two websites. 
To this end, we construct a {\em primary} dataset of 28,836,775 users
who have at least two plaintext passwords (61,552,446
passwords in total). Our analysis in the paper will focus on this
primary dataset. Note that users outside of the primary dataset are
not necessarily risk-free: they might still have accounts in services that we didn't cover.


\para{Ethic Guidelines.} Our work involves analyzing leaked datasets
that contain sensitive information. We have worked closely with our local
IRB and obtained the approval for our research. Our study is 
motivated by the following considerations. First, we only analyze datasets that are already publicly
available. Analyzing such data does not add additional risks other
than what already exist. Second, these datasets
are also publicly available to potential attackers. 
Failure to include the data for research may give attackers an
advantage over researchers that work on defensive
techniques. In the past decades, leaked password datasets have been
extensively used in academic
research~\cite{Ur:2015:,DasBCBW14,pars15,CarnavaletM14,VerasCT14,Wang:2016:,info16:}
to develop security mechanisms to protect users in the long run.

\begin{figure*}[t]
 \centering
\begin{minipage}{0.99\textwidth}
	\includegraphics[width=1\textwidth]{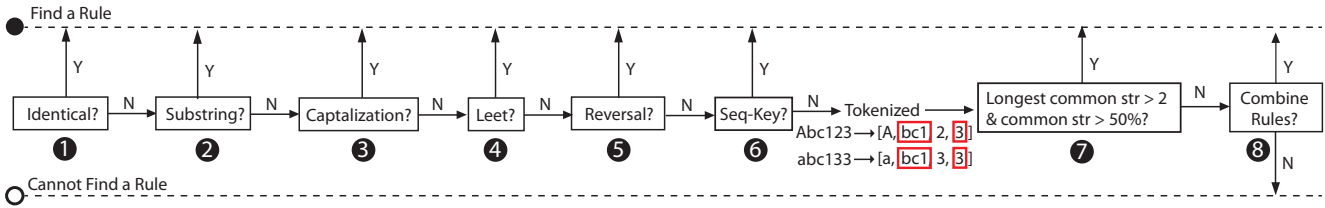}
	\vspace{-0.25in}
	\caption{The workflow to measure a user's password transformation
          patterns.}
	\label{fig:flow}
\end{minipage}
\vspace{-0.1in}
\end{figure*}

\section{Password Reuse Across Services}
\label{sec:reuse}
We start by analyzing how often users reuse {\em exactly the same}
password for different services. Out of the 28.8 million users in the
primary dataset, we extract 37,301,406 password pairs where both
passwords are from the same user. If a user has more than two
passwords, then all possible pairs are considered ({\em e.g.}, 4
passwords means 6 pairs). We find that 34.3\% of the pairs are
identical pairs. At the user level, 38\% of the users (10.9 million)
have at least one identical pair, indicating that the user
sets the same password for different services. This ratio is slightly lower than
the self-reported result (51\%) from a user study~\cite{DasBCBW14}.



Next, we are curious whether users with more passwords are more likely to
reuse the same password. The intuition is that it is difficult to
memorize many completely different passwords. Our result in
Table~\ref{tab:pwreuse} shows that this hypothesis
is true for users with less than 5 passwords. 
However, the trend is reversed for users
with even more passwords. A careful examination
shows that users with more passwords are more likely to ``modify''
an existing password for new services. Based on the results in
\S\ref{sec:rule}, users with more than 4 passwords have a higher
chance of modifying existing passwords (64.0\%)
compared with the overall ratio (21.0\%).


As a case study, we specifically analyze the passwords for ``email''
services. Email accounts are sensitive due to the fact that emails can be used to
reset the password for various online services. Many online
accounts will be in danger if the user's email account is compromised. 
As shown in Table~\ref{tab:cat}, we have 3
leaked email datasets from {\tt Gmail}, {\tt Mail.ru} and
{\tt Yandex}. We identify 4,033,847 password
pairs (involving 3 million users) that contain an email password. We find
that 2,440,232 of the pairs are identical, which yields a much higher
reuse ratio (60.4\%) than the overall ratio (34.3\%). This indicates
users are more likely to use their email password for another
service, a practice that incurs serious risks.

\begin{table}[t]
\centering
\begin{small}
\begin{tabu}{l|l|r}
\tabucline[1.1pt]{-}
\#PWs per User & \# of Users & \% of Users w/ PW Reuse \\
\hline 
2	& 25,515,516  & 34.6\% \\
3	& 2,877,322 & 63.4\% \\
4	& 370,990	& {\bf 78.8\%} \\
5	& 54,258	& 74.6\% \\
6	& 11,112	& 51.5\% \\
7	& 3,701	& 29.6\% \\
$\geq$ 8   & 3,876       & 22.6\% \\
\tabucline[1.1pt]{-}
\end{tabu}
\vspace{-0.12in}
\caption{Password reuse rate vs. \# of  passwords per user. }
\label{tab:pwreuse}
\end{small}
\vspace{-0.12in}
\end{table}

\section{Password Transformation}
\label{sec:rule}
In addition to reusing the same password, users may also modify an
existing password when signing up for a new service. 
Given a password pair of the same user, our goal is to infer the ``transformation rule''
(it there is one) that the user follows to modify the password. Then, we
seek to understand how much the transformation patterns differ across
users from different demographics. 




\subsection{Transformation Rules}
Our measurement workflow is shown in Figure~\ref{fig:flow}. In total,
we construct 8 rules for password transformation based on our manual
examinations of 1000 random password pairs and results from prior
studies~\cite{DasBCBW14,Zhang:2010:,7487945}. We test these rules against the
password pairs in the {\em primary dataset}, and the results are shown
in Table~\ref{tab:rulefound}. 


The majority of the password pairs (55.6\%) can be explained by one of
the transformation rules. To translate the numbers to the user level, 38\% of the users have reused the
same password at least once, and 21\% of the users have once
modified an existing password to create a new one. Collectively, these
users count for 52\%. The rest 48\% of the users are likely
to create a new password from scratch for each service. 
Below, we discuss each rule in details and further analyze the
unmatched passwords.

\begin{table}[t]
\begin{small}
\centering
\begin{tabu}{l|l|l}
\tabucline[1.1pt]{-}
Rule & \# Pairs of Passwords & Ratio (\%) \\ 
\hline
 \ding{202}. Identical & 12,780,722 & 34.3\%\\
 \ding{203}. Substring & 3,748,258 & 10.0\% \\
 \ding{204}. Capitalization & 478,233 & 1.3\% \\
 \ding{205}. Leet & 93,418 & 0.3\% \\
 \ding{206}. Reversal & 5,938 & < 0.1\% \\
 \ding{207}. Sequential keys & 12,118 & < 0.1\% \\
 \ding{208}. Common Substring & 2,103,888 & 5.7\% \\
 \ding{189}. Combination of Rules & 754,393 & 2.0\%\\
\hline 
Can Not Find A Rule & 17,324,438 & 46.4\% \\
\hline 
 Total & 37,301,406 &100\% \\ 
\tabucline[1.1pt]{-}
\end{tabu}
\vspace{-0.07in}
\caption{Distribution of password transformation rules.}
 \vspace{-0.2in}
\label{tab:rulefound}
\end{small}
\end{table}

\para{Identical.} The most common rule is reusing the same password
(12 million password pairs, 34.3\%). 

\para{Substring.} This rule indicates that one password is a substring
of the other one ({\em e.g.}, {\tt abc} and {\tt abc12}). This rule
matches 3.7 million password pairs (10\%), indicating that users have
inserted/deleted a string to/from an existing password to make a
new one. As shown in Table~\ref{tab:substring}, most
insertions/deletions happened at the tail (87.2\%). Most inserted/deleted
strings are pure digits (74\%) and short (1--2 characters), {\em
  e.g.}, ``{\tt 1}'', ``{\tt 2}'', and ``{\tt 12}''.  

\para{Capitalization.} Users may simply capitalize certain letters in
a password. Even though the ratio of matched pairs is not high (1.3\%), the
absolute number is still significant (478,233 pairs). We observe
that users commonly capitalize letters in the beginning of the password
(73\%), particularly the first letter (68.6\%). 

\para{Leet.} 93,418 password pairs match the leet
rule (0.3\%)~\cite{leet}. Leet transformation refers to replacing certain characters
with other similar-looking ones. 
Our analysis shows the top 10 most common transformations are: {\tt
  0}\ding{214}{\tt o}, {\tt 1}\ding{214}{\tt i}, {\tt 3}\ding{214}{\tt e},
{\tt 4}\ding{214}{\tt a}, {\tt 1}\ding{214}{\tt !}, {\tt 1}\ding{214}{\tt l}, {\tt
5}\ding{214}{\tt s}, {\tt @}\ding{214}{\tt a}, {\tt 9}\ding{214}{\tt 6}, and {\tt
\$}\ding{214}{\tt s}. These 10 transformations already cover 96.6\% of
the leet pairs. 

\begin{table}[t]
\begin{small}
\centering
\begin{tabu}{lr|lr}
\tabucline[1.1pt]{-}
Insert/Delete Position & Ratio &Inserted/Deleted Length & Ratio \\ 
\hline
Tail  & 87.2\% & 1 & 48.3\% \\
Head & 11.0\% &2 & 28.0\%\\
Both Ends & 1.8\% & 3+ & 23.7\% \\
\tabucline[1.1pt]{-}
Insert/Delete Type & Ratio & Top Inserted/Deleted Str. & Ratio \\
\hline
Digit & 74.0\% &  ``{\tt 1}'' & 24.2\%\\
Letter & 17.8\% &  ``{\tt 2}'' & 4.0\%\\
Combined & 4.5 \% &  ``{\tt 12}'' & 2.1\%\\
Special Char & 3.7\% &  ``{\tt 123}'' & 1.9\%\\
\tabucline[1.1pt]{-}
\end{tabu}
\vspace{-0.07in}
\caption{Substring rule: insertion/deletion patterns.}
 \vspace{-0.05in}
\label{tab:substring}
\end{small}
\end{table}

\begin{figure*}
\centering    
    \subfigure[Profession]{
      \includegraphics[width=0.32\textwidth]{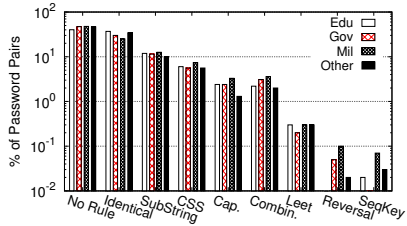}
       \vspace{-0.05in}
      \label{fig:p1}
      \vspace{-0.08in}
    }
    \hfill
    \subfigure[Country]{
      \includegraphics[width=0.64\textwidth]{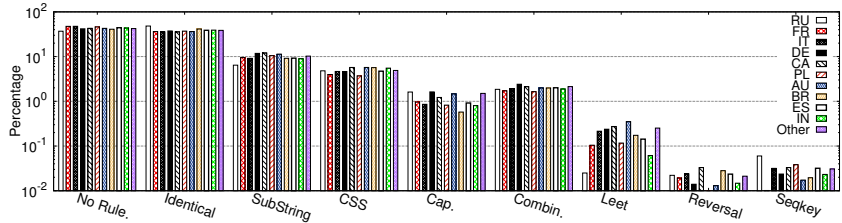}
     \vspace{-0.05in}
      \label{fig:p2}
      \vspace{-0.08in}
    }
\vspace{-0.05in}
  \caption{Distribution of password transformation rules for users
    of different professions and countries.  
  }
\vspace{-0.05in}
  \label{fig:pattern}
\end{figure*}

\para{Reversal.} Reversal rule is rarely used (5938
pairs, $<$0.1\%), which means reversing the order of the
characters in a password, {\em e.g.}, {\tt abcd}\ding{214}{\tt
  dcba}. Intuitively, reversed password is hard to memorize. 

\para{Sequential Keys.} Sequential keys include alphabetically-ordered letters ({\tt abcd}), sequential numbers ({\tt
  1234}) and adjacent keys on the keyboard ({\tt qwert}, {\tt asdfg}, {\tt
  !@\#\$\%}). The matched pairs ({\em i.e.}, both passwords are
sequential keys) are also below 0.1\%. 



\begin{table}[t]
\begin{small}
\centering
\begin{tabu}{lr|lr}
\tabucline[1.1pt]{-}
Longest Comm. Substring & Ratio & Transformation Rules & Ratio \\ 
\hline
Letter & 63.8\% & Substitution &84.7\% \\
Digit & 22.0\% & Insertion/Deletion & 32.4\% \\
Combined & 13.7\% & Capitalization & 3.2\% \\
Special Char & 0.5\% & Switching Order & 2.2\% \\
\tabucline[1.1pt]{-}
\end{tabu}
\vspace{-0.07in}
\caption{Common substring rule: longest common
  substring and transformation patterns.}
 \vspace{-0.15in}
\label{tab:comm}
\end{small}
\end{table}

\para{Common Substrings.} When a user modifies an existing
password to create a new one, we assume the majority of the password
remains the same. As shown in Figure~\ref{fig:flow}, we extract the
longest common substrings from the two passwords to learn how they transform
the rest parts. To avoid accidental character overlaps, we require the longest common
string to be $>$2 characters, and all the common
substrings should cover $>$50\% characters of a password
({\em i.e.}, the majority).

This rule matches 2.1 million password pairs (5.7\%). 
To make sure the thresholds make sense, we manually examine a random
sample of 1000 matched pairs. 44 pairs look like to have accidental
overlaps, which projects a false positive rate of 4.4\%. We can
tolerate false negatives for now since we have one more rule left. 
Based on the false positive rate, we estimate that the common substring rule should
count for at least 5.4\% of all password pairs.

Table~\ref{tab:comm} shows that the longest common substrings are often
pure letters (63.8\%) or pure digits (22\%). 56.7\% of the pure-letter
strings are English words/names (based on NLTK
corpus~\cite{Bird:2009:NLP:1717171}). Table~\ref{tab:comm} also shows
the typical transformations. Note that one password pair
may have multiple transformations (total exceeds 100\%).


\begin{table}[t]
\begin{small}
\centering
\begin{tabu}{lr|lr}
\tabucline[1.1pt]{-}
Rule Combination & Ratio & Rule Combination & Ratio \\ 
\hline
Capitalization+Substring  & 26.2\% &Reversal+CSS & 6.1\% \\
Leet+CSS & 21.8\% & Leet+SubString& 5.6\% \\
Seqkey+CSS & 13.2\% & Seqkey+SubString & 4.2\% \\
Reversal+Leet+CSS  & 7.1\% &Seqkey+Leet+CSS & 2.9\% \\
Capitalization+CSS & 6.2\%  & Others & 6.8\% \\
\tabucline[1.1pt]{-}
\end{tabu}
 \vspace{-0.07in}
\caption{Rule combinations (CSS: Common SubString).}
 \vspace{-0.05in}
\label{tab:comb}
\end{small}
\end{table}

\para{Combination of Rules.} As a final step, we combine possible rules
to find a match. Note that rule3--6 modify the characters (or the
sequence) in a password, while rule2 and rule7 operate on substrings. 
Our approach is to use a combination of rule3--6 to modify the
password first, and then test if rule2 or rule7 can declare a match.
In this way, we further matched another 754,000+ pairs
(2.0\%). Table~\ref{tab:comb} shows the most common ways of combining rules. 

\para{Unmatched Password Pairs.} After testing all the above rules, there
are 46.4\% of password pairs remain unmatched. To make sure we did not
miss any major rules, we randomly sample 1000 unmatched pairs for
manual examination. We did not find any password pair that still
exhibited a meaningful transformation. We regard the 46.4\% of
password pairs as the result of users ``making new passwords from scratch''.



\subsection{Impact of User Demographics}
Next, we seek to understand how much the transformation patterns differ across
different user demographics. We infer user demographics from their email addresses. 

\para{Profession.} Certain email domains are exclusive to people of
special organizations. For example, ``{\tt .edu}'' is limited to higher educational
institutions, ``{\tt .mil} '' is exclusively to military, and   ``{\tt
  .gov}'' represents government agencies. In total, we identify
128,036 users from educational institutions, 7,376 users from military,
and 3,384 users from the government. As shown in Figure~\ref{fig:p1},
we find that their password transformation patterns are 
surprisingly consistently: about 30\% password pairs are identical,
followed by those that apply the substring rule (about 10\%), common
substring rule (about 5\%) and capitalization rule (about 3\%). Rules
such as leet, reversal and sequential key are consistently below 1\%.

\para{Country.} Similar results are observed in Figure~\ref{fig:p2},
where we divide users based on their countries. More specifically, we
identify email domains that contain a country code ({\em e.g.}, ``{\tt
  .ru}'' stands for Russia). This returns 233 country codes and
5,892,528 users. In Figure~\ref{fig:p2}, we plot the distributions of the transformation rules for
the top 10 countries (counting for 90.5\% of the users with a
country code). Again, the transformation patterns are very similar for
users from different countries. 

Our result demonstrates a high-level of consistency (low
variance) for password transformation patterns across different user
populations. This, however, could make the attacker's job
easier. Even using a small dataset, it is possible for the attacker to
learn the basic transformation patterns that apply to broader user
populations. In the next section, we develop a {\em
  training-based} algorithm to validate this hypothesis.

\section{Password Guessing}
\label{sec:guess}
Based on the measurement results, we then evaluate the security risks
introduce by users modifying an existing password for different
services. We perform password guessing experiments using a {\em
  training-based} algorithm to answer two key
questions: First, how quickly can attackers guess a modified password
based on a known one? Second, given the low variance of
password transformation patterns, can attackers use a small training
data ({\em e.g.}, 0.1\%) to achieve effective guessing? 

\subsection{Guessing Algorithm}
We build a new password guessing algorithm by addressing the
weaknesses in DBCBW~\cite{DasBCBW14}. DBCBW is a popular algorithm to
guess a target user's password by transforming a known password of the
same user. DBCBW's design goal is simplicity, but has two weaknesses: First, due
to the lack of training data, the algorithm uses hand-crafted
transformation rules. Second, it applies these rules in a
{\em fixed order}, which may not be optimal for individual
passwords. For example, ``{\tt l0ve}'' should try the leet rule first ({\tt
  0}\ding{213}{\tt o}), even though the substring rule is overall more
popular.

Our algorithm overcomes these drawbacks by introducing a training phase. Using
ground-truth password pairs as the training data, we learn two
things: (1) the transformation procedure for each rule, and (2) a
model to customize the ordering of the rules for each password. 

\para{Training: Transformation Procedures. } For each rule $R_i$, we seek
to learn a list of password transformations $T_i=[t_{i1}, t_{i2}, ... t_{iN_i}]$ where
$t_{ij}$ represent one transformation under this rule. $T_i$ is sorted by
the frequency of each transformation's appearance in the training
dataset. During password guessing, we will test each transformation
{\em independently}. For example, in the ``substring rule'', $t$ is characterized by
{\em <insert/delete><position><string>}. In ``capitalization
rule'', $t$ is characterized by {\em <position><\#chars>}. In a
similar way, we learn the transformation list $T$ for ``leet'', ``sequential keys''
and ``reversal''. 

Common substring rule is special. During training, we learn the
sorted transformation list (insert, delete, replace, substitute, switch orders). However, when applying
the transformation to a given password, we need to first split the password
to detect potential common substrings. In our design, we test 3 types of candidate:
(1) substrings of pure digits/letters/special characters, (2) English words/names, and (3) popular common substrings in
the {\em training data}. For the ``combined rule'', $T$ is a sorted list of
rule-combinations where each rule-combination has a sorted list of
transformations to be tested. 



\begin{figure*}
\begin{minipage}{0.68\textwidth}
\centering    
    \subfigure[5000 Guesses]{
      \includegraphics[width=0.47\textwidth]{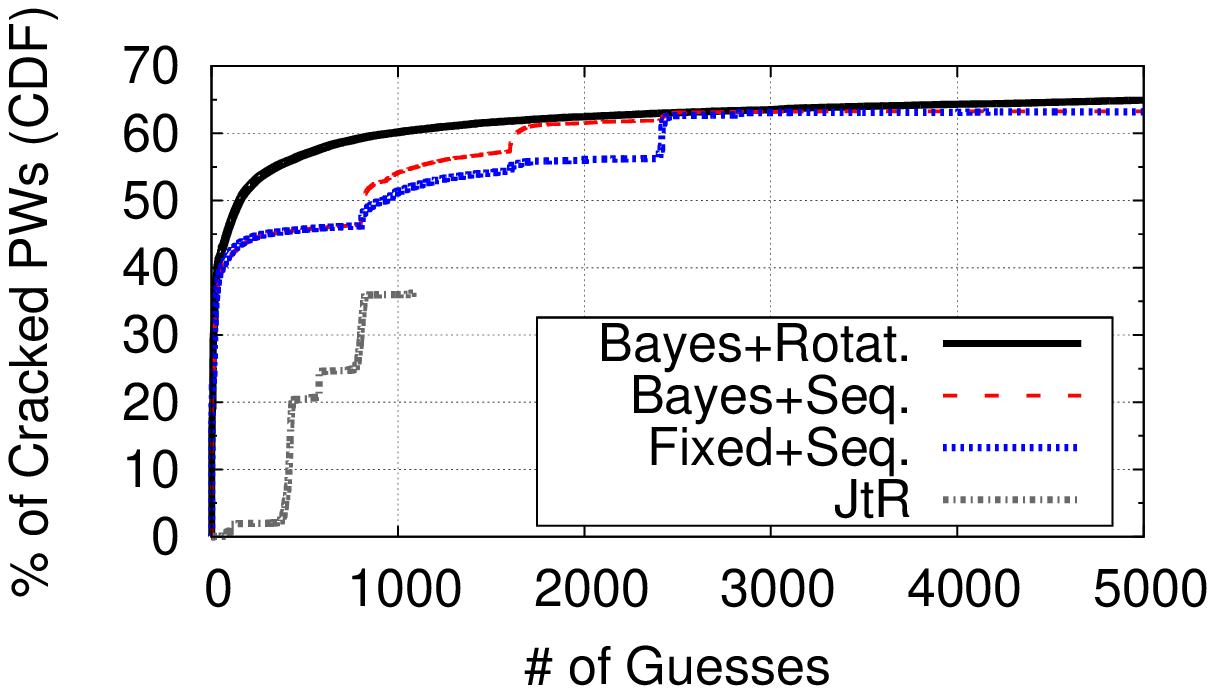}
       \vspace{-0.05in}
      \label{fig:g1}
      \vspace{-0.08in}
    }
    \hfill
    \subfigure[10 Guesses]{
      \includegraphics[width=0.47\textwidth]{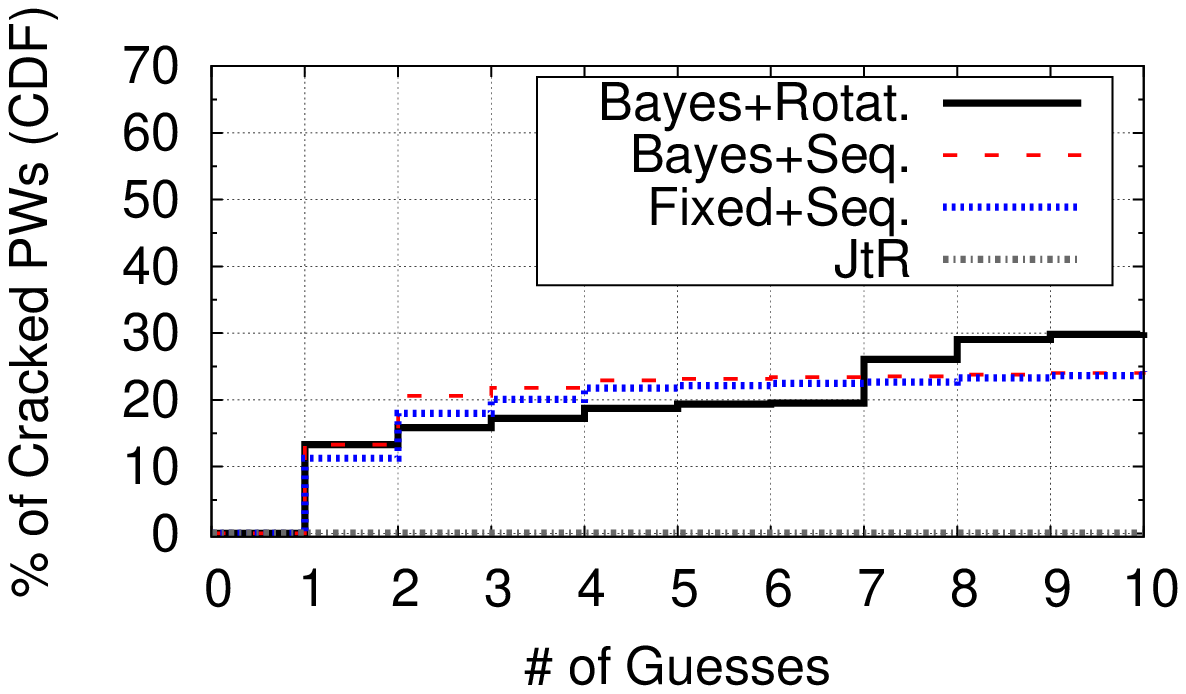}
     \vspace{-0.05in}
      \label{fig:g2}
      \vspace{-0.08in}
    }
\vspace{-0.16in}
  \caption{Password guessing with 50\% of the data for training. }
\vspace{-0.05in}
  \label{fig:guess}
\end{minipage}
\vspace{-0.03in} 
\hfill 
\begin{minipage}{0.31\textwidth}
 \centering
	\includegraphics[width=1\textwidth]{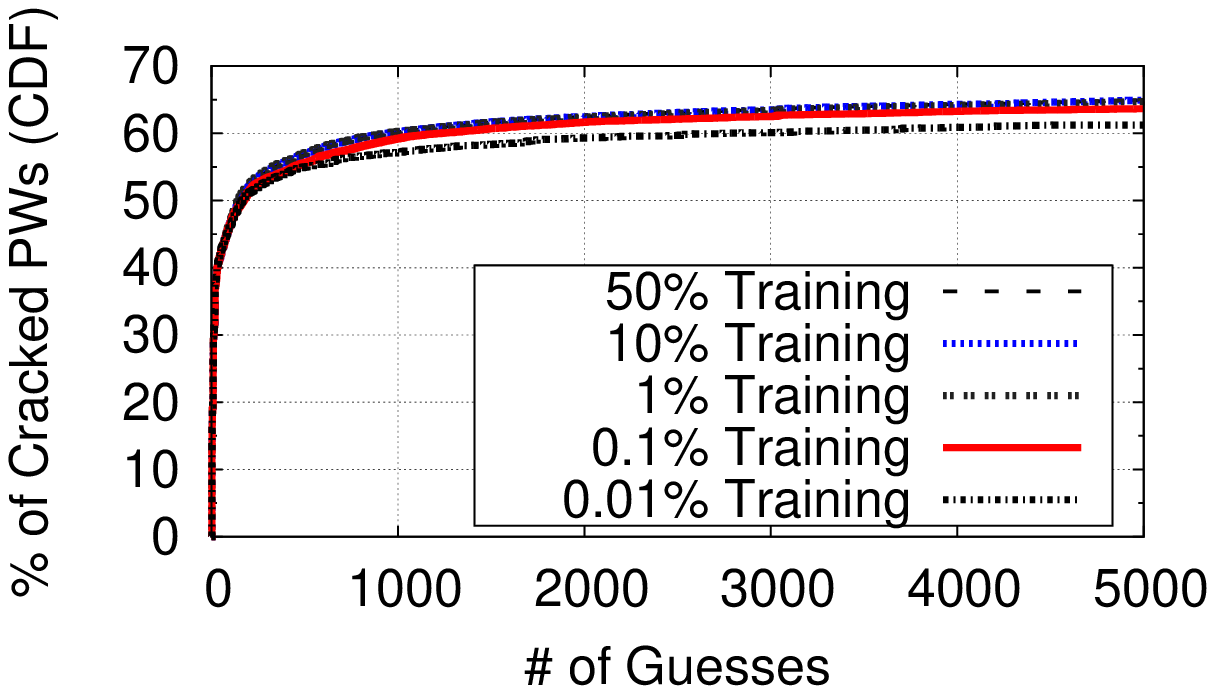}
	\vspace{-0.27in}
	\caption{Password guessing with different training data sizes.}
	\label{fig:train}
\end{minipage}
\vspace{-0.03in} 
\end{figure*}

\para{Training: Rule Ordering.} For a given password, we
learn which rule should be applied first using a Bayesian model. 
We treat this as a multiple-class classification problem. Given a
password, we train a model to estimate the likelihood that the
password can be transformed by each rule. To achieve a quick training,
we choose the Naive Bayes classifier (multinomial model)~\cite{MB2008},
which produces the {\em probability} that a data point
(password) belong to a class (rule). Based on the probability, we
customize the ordering of the rules for this password. 
Table~\ref{tab:feature} shows the 18 features used in the Bayesian model. 

\para{Password Guessing.} For a given password pair $(pw_1, pw_2)$,
we test how many attempts are needed to guess $pw_2$ by transforming
a known $pw_1$. 
We first use the Bayesian model to generate a customized order of rules for
$pw_1$. Following the ordered rule list, we have two options for guessing:  
\begin{packed_itemize}
\item {\em Sequential}: testing one rule at a time. After testing
  all the transformations under a rule, we move to the next
  rule. Since certain rules have a significantly longer list than
  others, we set a threshold $M$ as the maximum number of
  guesses under each rule ($M=800$ for our experiment).  
\item {\em Rotational:} testing one rule and one transformation at
  a time. After testing one transformation under a rule, we move to
  the next rule to test another transformation. We rotate to test each
  rule for just one guess. 


\end{packed_itemize}
Sequential guessing requires a higher accuracy of the predicted
order. If the predicted order is wrong, it will waste many guesses on
the wrong rule before moving on. Rotational method is more tolerable
to the prediction errors



\begin{table}[t]
\begin{small}
\centering
\begin{tabu}{l}
\tabucline[1.1pt]{-}
{\bf 18 Features Extracted from a Password} \\ 
\hline
PW (password) length, \# Lowercase letters, \# Uppercase letters,  \#
Digits, \\
\# Special chars, Letter-only pw?, Digit-only pw?, \#
Repeated chars, \\
Max \# consec. letters, Max \# consec. digits,
Max \# sequential keys, \\
Englishword-only pw?, \# Consec. digits (head), \# Consec. digits (tail), \\
\# Consec. letters (head), \# Consec. letters (tail), \\
\# Consec. special-chars (head), \# Consec. special-chars (tail) \\
\tabucline[1.1pt]{-}
\end{tabu}
 \vspace{-0.1in}
\caption{Feature list of the Bayesian model. }
  \vspace{-0.15in}
\label{tab:feature}
\end{small}
\end{table}

\para{Baselines.}
We use two baselines for comparison. First, instead of customizing the
order for each password, we apply the rules with {\em a fixed order} for ``sequential
guessing'' (similar to DBCBW). The fixed order is based on the overall rule popularity in
the training data. Our second baseline is a popular off-the-shelf password
cracking tool John the Ripper (JtR)~\cite{JtR}. We use the ``single'' mode
of JtR and keep the default setting. Given a password, JtR applies a list of mangling rules to
transform the password. The guessing stops when all the mangling rules are exhaustively tested. 



\subsection{Password Guessing Results}
We use the proposed algorithm to evaluate the risks of modified
passwords. For this experiment, we exclude identical password pairs
(34.3\%) since they only take one guess, and 46.4\% of pairs that
did not match a rule ({\em i.e.}, new passwords created from
scratch). This leaves us 7,196,242 password pairs that represent password
modifications ({\em exp dataset}). We conduct two experiments. First,
we split the {\em exp dataset} to use 50\% for training and the other 50\%
for testing. Second, to validate the ``low-variance''
assumption we try to use much smaller training data. 

\para{Training on 50\% of the Data.} During password guessing, we test 
both directions for each password pair ($pw_1$\ding{213}$pw_2$ and
$pw_2$\ding{213}$pw_1$), which doubles the testing data.  
As shown in Figure~\ref{fig:guess}, our best algorithm guessed 46.5\% of the
passwords within just 100 attempts. 
Figure~\ref{fig:g2} shows that 10 guesses already cracked 30\% of the passwords. In
comparison, the JtR baseline almost got nothing in the
first 10 attempts and exhausted all the mangling rules after 1081
guesses. Since we evaluate an online-guessing scenario, we stopped our
algorithm after 5000 guesses for each password.\footnote{Our experiment
  shows that 50,000 guesses can crack 70\%.}

Comparing different algorithms, we show that the Bayesian
model outperforms the fixed ordering method. This confirms the
benefits to prioritize the more likely
rules for each password. In addition, rotational guessing is
better than sequential guessing. Sequential
guessing has a clear stair-step increase of the hit rate after
switching to a new rule. This indicates that the first few
transformations under each rule are the most effective ones. 
Sequential guessing's advantage is in the first 5 guesses
(Figure~\ref{fig:g2}) --- if the Bayesian prediction
is correct, sticking to the right rule helped to guess the password
quicker. Rotational guessing has an overall
better performance by switching the rules more frequently.

\para{Using Smaller Training Data.} To validate the low-variance
assumption of the transformation patterns, we use even smaller
datasets to train our algorithm (Bayesian+rotational). We
vary the size of the training data from 0.01\% to 10\% of the {\em exp
  dataset}. To be consistent, we use the same 50\% as the testing data
(training and testing data has no overlap). As shown in
Figure~\ref{fig:train}, the 0.1\%-training curve is still overlapped
with the 50\%-curve, suggesting that extremely
small training data can achieve a comparable performance. 
This confirms the low-variance in password transformation patterns. 
Intuitively, users modify a password for the ease of
remembering. This is likely to introduce easy-to-predict passwords. 

To measure the number of vulnerable password pairs, we use
the 0.1\%-trained model to guess the rest 99.9\% of the
password pairs. Since we guess both directions, the testing data
essentially has 14 million passwords. Within 10 attempts, we guessed 30\% (4.2
million passwords) --- 3.8 million {\em password pairs} are cracked for at
least one direction. Together with the identical password pairs (12.8
million), over 16.6 million pairs can be cracked within 10 attempts. 




\para{Cracking the Remaining Hashes.} Finally, we perform a quick
experiment on the uncracked hashes in Section~\ref{sec:data}. In
total, we have 6,218,778 password pairs where one password is an uncracked hash, and
the other one is in plaintext. Our algorithm successfully recovered
939,400 (15.1\%) of the hashes within 5000 attempts, which
demonstrates the value of our algorithm over existing offline cracking
tools. As a future work, we plan to further test our algorithm on {\em
  salted} password hashes.

\section{Discussion}
After analyzing 28.8 million users' passwords across 107 services, we
find that a majority of users have reused the same password or slightly modified
an existing password for different services. Password modification
patterns are highly consistent across various user populations,
allowing attackers to crack massive passwords {\em online} with
minimal training. 

Moving forward, the challenge is how to effectively mitigate the
threat after a service is breached. Given the high reuse
rate of passwords, it is necessary to immediately notify users to
reset the password, not only for the breached service but also other services with a similar
password. The question is who should play the role to notify users,
given that not all the breached services would immediately disclose
the incident or contact users~\cite{new16, new15}. In addition,
during password reset, it is critical to make sure users don't modify
the already-leaked password to create the new one. A better practice is to use
password managers ({\em e.g.}, 1Password) to set unique and
complex passwords for each service without the need to memorize
them. Finally, our result shows a concerning
high ratio of {\em email password reuse}. We argue that more specific warnings should be given
to users to avoid reusing the email password when signing up for a service.

\newpage
\balance
\begin{small}
\bibliographystyle{acm}
\bibliography{astro}
\end{small}

\end{document}